\documentclass{webofc}

\usepackage[varg]{txfonts}   
\usepackage{hyperref}
\usepackage{url}
\hypersetup{colorlinks=true,citecolor=blue,urlcolor=blue,linkcolor=blue}
\begin{document}
\title{Theory Summary}
%
%

\author{\firstname{Hannah} \lastname{Elfner}\inst{1,2,3,4}\fnsep\thanks{\email{h.elfner@gsi.de}}}

\institute{GSI Helmholtzzentrum f\"ur Schwerionenforschung,  64291
		Darmstadt, Germany
		\and
           Goethe University Frankfurt, Department of Physics, Institute for Theoretical Physics, 60438 Frankfurt, Germany
           \and
           Frankfurt Institute for Advanced Studies,  60438
		Frankfurt am Main, Germany
           \and
           Helmholtz Research Academy Hesse for FAIR (HFHF), GSI Helmholtz Center,
		Campus Frankfurt,  60438 Frankfurt am Main, Germany
            }

\abstract{This is the write-up of the summary talk on theory activities in the heavy-ion community as presented at Quark Matter 2025. It contains a (biased) selection of results from the parallel program of the conference. The progress is reported for three different areas: Bulk dynamics and initial state; jets, heavy flavor and electromagnetic probes and QCD phase diagram and observables. The idea is to convey the major achievements in the field and give a perspective for future directions. 
}
\maketitle
\section{Introduction - setting the stage}
\label{intro}
Quark Matter 2025 has been the largest conference in this series so far with close to 1000 participants. In this summary, I chose to focus on the parallel program, which featured 100 theory contribution. Just dividing the 30 minutes of the summary talk by 1500 minutes parallel program, I can cover about 2 \% of the content here. This results in a biased selection of topics and specific results that are mentioned in the following.

In general, the interaction of theory and experiment is crucial for progress in the field, since new theory developments are often inspired by experimental discoveries and experimental efforts are usually guided by theory predictions. Nevertheless, as it was the charge from the organizers, I concentrate here on the theory developments and the experimental summary is covered in another contribution. 

Besides the actual scientific achievements what I have noticed is that there are more and more structured larger collaborations of theorists working on relativistic heavy-ion collisions, sometimes also joining with expertise from neighboring fields. This collaborative effort combined with the new ideas of individuals lays the foundation to make progress in an increasingly complex environment. In addition, the open sharing of data and software which happens more frequently helps to address complicated detailed issues relying on all available expertise.  

\section{Bulk dynamics and initial state}
\label{sec-bulk}
All physics knowledge that we extract from collisions of highly accelerated ions about properties of QCD (quantum chromodynamics) matter requires sophisticated theoretical modeling. It is an important advance that the state-of-the-art dynamical approaches are globally applied to understand many different observables at the same time. While the 'standard model' based on non-equilibrium initial conditions, relativistic viscous hydrodynamics and hadronic transport is well established, the interfaces between the different stages still imply considerable uncertainties. The initial evolution and approach to equilibrium is still under heavy investigations as well as the existing uncertainties in the final stage decoupling due to the different prescriptions for handling the viscous corrections. In \cite{Gavassino:2021owo} the limits of hydrodynamics are thoroughly addressed and it has been shown, that our numerical codes sometimes solve the equations properly even in acausal regimes. 

Even with the main ingredients being established we are facing a many-parameter-multi-observables challenge that is addressed with Bayesian statistical analysis that are commonly employed as for example by the JETSCAPE collaboration in \cite{Mankolli}. A new avenue is to replace the costly evolution by neural network calculations and \cite{Auvinen} showed that this is indeed possible without loosing too much information. As a next step, it would be nice to proceed to meta-analysis where the results of different Bayesian analysis are combined to one conclusion. 

The main advancements that have been presented at this conference address different features of the initial state and its evolution. The recently established connection to the nuclear structure physics community is interesting, in particular in view of the oxygen-oxygen and neon runs at the LHC this summer. Analyzing the anisotropic flow coefficients in heavy-ion collisions offers a more direct 'view' of the initial state deformations than nuclear structure experiments. A Bayesian analysis of PbPb and XeXe data revealed clear evidence of the triaxial structure of xenon \cite{van der Schee}. Clear differences are expected from the 'bowling pin'  structure of neon compared to round nuclei \cite{Giacalone:2024ixe}. This proves that at the current level of accuracy the shape details have to be taken into account for our modeling of the bulk dynamics. 

Many calculations have been and are performed assuming boost invariance reducing the actual calculation to two spatial dimensions in the transverse plane, but longitudinal structures deserve some attention. In particular, it is nice to see new fast Monte Carlo generators like MCDipper including finite charge deposition \cite{Garcia-Montero:2025bpn}. Also the EKRT approach is now extended to three dimension and available for fast generation of initial state profiles \cite{Kuha:2024kmq}. I think one interesting question to be resolved is to understand the longitudinal structures differentiating between a string-based approach or a saturation-based glasma. 

Concerning the initial non-equilibrium evolution the work in \cite{Mantysaari:2025tcg} relies on the real-time evolution of the gauge fields including the full JIMWLK evolution for Bjorken x dependence. Those calculations are constrained to deep inelastic scattering results and then applied to heavy-ion collisions allowing for a comprehensive understanding of different collision systems. As a second example, QCD kinetic theory calculations can be expedited by machine learning techniques. In \cite{BarreraCabodevila:2025ogv} the collision kernel is approximated by a neural network and the density evolution is well reproduced. Pushing such efforts further will lead to an improved understanding of the initial non-equilibrium evolution and its matching to hydrodynamics. In \cite{Sousa:2024msh} it has been shown that finite momentum space information improves the prediction of final state correlations significantly in particular in small systems.

\section{Jets, heavy flavor and electromagnetic probes}
\label{sec-hard}

Hard probes have to be embedded in a realistic dynamic medium evolution to gain an overall understanding of QCD properties. One example for a semi-analytical calculation with vacuum-like evolution and medium-induced emissions has been embedded in a hybrid approach. A large set of jet observables can be understood in one single framework \cite{Mehtar-Tani:2021fud,Takacs:2021bpv,Mehtar-Tani:2024jtd}. Interestingly, this approach predicts small quenching in oxygen-oxygen collisions while a significant $v_2$ coefficient is predicted, a potential explanation why the quenching has not been observed in small systems yet. 

All jet and jet substructure observables need to be scrutinized for bias effects and the sensitivity to the physics of interest. In \cite{Andres:2024hdd,Andres:2024pyz} the energy-energy correlators that are suggested as a new observable distinguishing the relevant scales have been corrected for energy loss effects. The remaining modification hints then at the medium interaction of interest. Full Monte Carlo simulations like \cite{JETSCAPE:2025rip} allow for understanding the origin of differences in the final observables related to the nature of the parton that initiated the shower. 

Complementary to the more large-scale simulations, there have been advances in understanding the in-medium jet shower evolution. Even before the splitting, the finite formation time of the antenna has an effect on the critical angle to resolve radiation, since the interactions with the medium introduce more dependencies on other kinematic quantities \cite{Abreu:2024wka}. In \cite{Pablos:2024muu} the coherence effect of recoils has been studied and it is suggested how to implement this effect to better understand jet-medium interaction. 

The long neglected influence of the initial non-equilibrium evolution on the energy loss of hard partons has been investigated in QCD kinetic theory where more kicks along the beam direction have been observed \cite{Lindenbauer}. Looking at the broadening of the quark jet in a Yang-Mills field glasma evolution the same effect has been found \cite{Lamas}. For heavy quarks the influence of the initial pre-equilibrium evolution seems to be rather limited as presented in \cite{Schenke}. The nuclear modification factor is slightly different, but the elliptic flow is insensitive. 

Detailed dynamical evolution models including event-by-event fluctuations are important as a background to understand heavy flavor observables. One of the main quantities of interest is the diffusion constant. In \cite{Sambataro:2024mkr} it has been shown that the momentum dependence of $D_s$ is crucial to describe observables.  When employing different approaches for hadronization, it has been shown that the sensitivity to the lattice input for the diffusion constant is maintained \cite{FuBass}. A really impressive new result is the heavy quark diffusion constant calculated from lattice QCD over a broad range of temperatures very close to the physical pion mass \cite{HotQCD:2025fbd}. These new results are now ready to be implemented in phenomenological calculations and provide one more direct link of QCD properties to experimental data. 

The screening of quarkonium bound states at high temperatures has been under investigation since many years. The newest results for the $q-\bar{q}$ potential indicate almost no change at finite temperature compared to the vacuum case \cite{Bazavov:2023dci}. The calculations of spectral functions have advanced to a level that allow the extraction of a width as a function of temperature. The results from \cite{Larsen} suggest a broadening which indicates that the bound states are destroyed by scatterings and not necessarily due to a screened potential. 

Electromagnetic probes are challenging to measure and only few theory contributions were devoted to this topic at this conference. In the light of the expectation of more high precision data from future experiments it is the right time to refine theoretical predictions. In \cite{Coquet:2023wjk} the dilepton anisotropy has been suggested as a measure providing access to the early time evolution. Global calculations such as \cite{Wu} within dynamic hybrid approaches that describe bulk properties and the emission of electromagnetic probes are important to gain a complete understanding. 

\section{QCD Phase diagram and observables}
\label{sec-qcd}
While the phase diagram along the temperature axis has been well established the finite density regime is of high interest. \cite{Pirelli} presented canonical lattice simulations that do not suffer from the sign problem to extend the equation of state to finite net baryon chemical potential. In \cite{Borsanyi:2024xrx} the deconfinement transition has been determined to be slightly above the chiral transition, which intuitively makes sense, the quarks reduce their masses first and then become deconfined. \cite{Shah:2024img} attempts to locate the critical endpoint by different extrapolations of lattice QCD data. Even though those methods might not be fully conclusive the result ends up in a similar region as current functional QCD calculations \cite{Rennecke}.

In \cite{STAR:2025zdq}  the STAR collaboration has released the final results on the net proton cumulants from the beam energy scan program at RHIC. Those results are not fully consistent with the baseline calculations without any critical behavior. To draw any further conclusions theory efforts are mandated that assess how much fluctuations are expected from a complete dynamical calculation including a phase transition and a critical endpoint. For example, \cite{Bzdak:2025rhp} points out that antiproton fluctuations may require a specific treatment. 

Fluid dynamics formulated in the density frame and enhanced by stochastic fluctuations might pave the way to the required calculations. \cite{Bhambure:2024axa} presented the dynamics in reduced dimensions agreeing with QCD kinetic theory, while \cite{Chattopadhyay:2024jlh,Chattopadhyay:2024bcv} presented full 3D box calculations with proper fluctuation dynamics. There were also presentations showing the progress on understanding the first order phase transition dynamics. In \cite{Karthein:2024zvs} trajectories were determined that show the expected features of a spinodal region. The spinodal decomposition has been dynamically simulated in \cite{Kapusta:2024nii}, where the corresponding bubble formation has been observed. 

At high chemical potential and lower temperatures neutron star mergers explore the QCD phase diagram. To connect the results for the equation of state of nuclear matter from heavy-ion collisions at low beam energies to constraints obtained from astrophysical observations is a promising avenue to arrive at a global understanding of dense strongly-interacting matter. The main difference between the two systems concerns the isospin which has to be taken into account. \cite{Mroczek:2024sfp} presented a new way to construct an equation of state at finite temperature, isospin and net baryon density incorporating neutron star constraints. The existence of hyperons in neutron stars has been questioned by the high mass neutron stars that need to be accommodated by the equation of state which is typically too soft, if it involves hyperons. Calculations by chiral effective field theory suggest a non-trivial hyperon-nucleon interaction that depends on density and momentum. Such a potential has been implemented in a transport approach and compared to hyperon flow measurements in heavy-ion collisions \cite{Jinno:2023xjr}. 

The measurement of finite polarization of $\Lambda$ hyperons as well as the spin alignment of vector mesons has sparked a lot of theoretical developments. Understanding vorticity, magnetic fields and spin dynamically evolving in heavy-ion collisions poses a major challenge in this highly relativistic environment. From the wealth of theory contributions on this topic at this conference, here are two examples of new results. In \cite{Giacalone:2025bgm} it is suggested that the nucleon fluctuations in the initial state potentially generate a non-zero polarization event by event which needs to be taken into account for the following dynamical evolution. Comparing the spin relaxation time to the usual shear relaxation indicates that scale separation can only be assumed under certain kinematic conditions \cite{Wagner:2024fhf}. Full numerical simulations are required to make contact with experimental data on polarization observables. Understanding the vortical dynamics of the system created in heavy-ion collisions will constrain the dynamic evolution and therefore help to better understand phase transition observables in the future. 

\section{Summary and future directions}
\label{sec-sum}

In the future, there are many exciting experimental campaigns on the horizon: CBM at FAIR will start data taking in 2029, EIC at BNL as well as ALICE3  and NA60+ at CERN are scheduled in the 2030's. Those experiments need to be accompanied by a reliable theory effort that provides predictions and understanding of the measurements. Further refining our standard model for the dynamical evolution requires consistently matching all stages, from initial and final non-equilibrium to the relativistic hydrodynamic evolution. 

Comprehensive approaches of hard and soft physics observables in one framework are confronted with detailed high precision measurements. Beyond the most active theory developments at this conference are the investigations concerning the initial non-equilibrium evolution, spin dynamics and polarization and the phase transition dynamics. To answer the legacy question of our field, namely "How much further can transport properties be constrained?" and "Is there a first order phase transition and implicitly a critical endpoint?" need joined forces. Open developments help in this respect and make the scientific effort sustainable. It is going to be exciting to exploit the new input from lattice and functional QCD to phenomenological approaches to strengthen the direct link between fundamental theory and experimental measurements. 

\section{Acknowledgements}
I would like to thank all speakers at the conference and colleagues who have pointed me to results of interest or shared their preliminary work with me prior to and at the conference. All discussions are greatly appreciated and have helped to bring this contribution together. 

%
%
%

\end{document}